\documentclass[twocolumn,english,showpacs,showkeys,reprint]{revtex4}
\usepackage[T1]{fontenc}
\usepackage[latin9]{inputenc}
\usepackage{array}
\usepackage{float}
\usepackage{textcomp}
\usepackage{amsmath}
\usepackage{amssymb}
\usepackage{graphicx}
\usepackage{setspace}

\makeatletter

\newcommand{\lyxmathsym}[1]{\ifmmode\begingroup\def\b@ld{bold}
  \text{\ifx\math@version\b@ld\bfseries\fi#1}\endgroup\else#1\fi}

\providecommand{\tabularnewline}{\\}

\@ifundefined{textcolor}{}
{%
 \definecolor{BLACK}{gray}{0}
 \definecolor{WHITE}{gray}{1}
 \definecolor{RED}{rgb}{1,0,0}
 \definecolor{GREEN}{rgb}{0,1,0}
 \definecolor{BLUE}{rgb}{0,0,1}
 \definecolor{CYAN}{cmyk}{1,0,0,0}
 \definecolor{MAGENTA}{cmyk}{0,1,0,0}
 \definecolor{YELLOW}{cmyk}{0,0,1,0}
}

\@ifundefined{definecolor}
 {\usepackage{color}}{}

\@ifundefined{textcolor}{}{%
 \definecolor{BLACK}{gray}{0}
 \definecolor{WHITE}{gray}{1}
 \definecolor{RED}{rgb}{1,0,0}
 \definecolor{GREEN}{rgb}{0,1,0}
 \definecolor{BLUE}{rgb}{0,0,1}
 \definecolor{CYAN}{cmyk}{1,0,0,0}
 \definecolor{MAGENTA}{cmyk}{0,1,0,0}
 \definecolor{YELLOW}{cmyk}{0,0,1,0}
 }

\usepackage{babel}
\usepackage{babel}

\makeatother

\usepackage{babel}
\begin{document}

\title{Co$_{2}$FeAl Heusler thin films grown on Si and MgO substrates:
annealing temperature effect}

\author{M. Belmeguenai %
\footnote{belmeguenai.mohamed@univ-paris13.fr%
}$^{1}$ , H. Tuzcuoglu$^{1}$ , M. S. Gabor%
\footnote{mihai.gabor@phys.utcluj.ro%
}$^{2}$ , T. Petrisor jr$^{2}$ , C. Tuisan$^{2,3}$ , F. Zighem$^{1}$
, S. M. Chérif$^{1}$ and P. Moch$^{1}$ }

\affiliation{$^{1}$ LSPM-CNRS, Université Paris 13, 99 avenue Jean-Baptiste Clément
93430 Villetaneuse, France }

\affiliation{$^{2}$ Center for Superconductivity, Spintronics and Surface Science,
Technical University of Cluj-Napoca, Str. Memorandumului No. 28 RO-400114
Cluj-Napoca, Romania}

\affiliation{$^{3}$ Institut Jean Lamour, CNRS, Université de Nancy, BP 70239,
F\textendash{} 54506 Vandoeuvre, France }
\begin{abstract}
10 nm and 50 nm Co$_{2}$FeAl (CFA) thin films have been deposited
on MgO(001) and Si(001) substrates by magnetron sputtering and annealed
at different temperatures. X-rays diffraction revealed polycrystalline
or epitaxial growth (according to the relation CFA(001){[}110{]}//MgO(001){[}100{]}
epitaxial relation), respectively for CFA films grown on a Si and
on a MgO substrate. For these later, the chemical order varies from
the $A2$ phase to the $B2$ phase when increasing the annealing temperature
($T_{a}$) while only the $A2$ disorder type has been observed for
CFA grown on Si. Microstrip ferromagnetic resonance (MS-FMR) measurements
revealed that the in-plane anisotropy results from the superposition
of a uniaxial and of a fourfold symmetry term for CFA grown on MgO
substrates. This fourfold anisotropy, which disappears completely
for samples grown on Si, is in accord with the crystal structure of
the samples. The fourfold anisotropy field decreases when increasing
$T_{a}$ while the uniaxial anisotropy field is nearly unaffected
by $T_{a}$ within the investigated range. The MS-FMR data also allow
for concluding that the gyromagnetic factor remains constant and that
the exchange stiffness constant increases with $T_{a}$. Finally,
the FMR linewidth decreases when increasing $T_{a}$, due to the enhancement
of the chemical order. We derive a very low intrinsic damping parameter
($1.1\times10^{-3}$ and $1.3\times10^{-3}$ for films of 50 nm thickness
annealed at 615$\lyxmathsym{\textdegree}C$ grown on MgO and on Si,
respectively).
\end{abstract}


\maketitle

\section{Introduction}

The future spintronic devices require an ideal spin-polarized electron
source, achievable by using a half-metallic Heusler alloys {[}1, 2{]},
having the composition $X_{2}YZ$ ($X$ being a transition metal element,
$Y$ being another transition metal element and $Z$ being a group
III, IV, or V element). These materials are expected to provide very
large magneto-resistive effects when used as magnetic electrodes in
magnetic tunnel junctions (MTJs) and in current-perpendicular-to-plane
(CPP) spin valves. They can be used as perfect spin filters and spin-injection
devices as alternative materials to ferromagnetic 3d metals. Therefore,
Co-based Heusler alloys, such as Co$_{2}$FeAl, have attracted much
research interest due to their large magnetic moment and their high
Curie temperature. Co$_{2}$FeAl (CFA) has a very high Curie temperature
(1000 K) and is theoretically predicted to have a half-metallic character
arising from its spin-split band structure. It can provide giant tunnelling
magneto-resistance (360\% at room temperature) {[}3, 4{]} when used
as an electrode in magnetic tunnel junctions, which makes CFA promising
for practical applications. However in such alloys, there is always
some degree of chemical disorder, which strongly influences many of
their physical properties. In reality, the totally ordered phase (L21)
is difficult to achieve and there is a variety of derived structural
types arising from atomic disorder in the occupation of the available
sites. When the X atoms occupy their assigned sites for the $L2{}_{1}$
phase, while the $Y$ and $Z$ atoms randomly share the other ones,
the $B2$ structure is obtained. The structure $A2$ corresponds to
a completely random occupation, by any $X$, $Y$ or $Z$ atom, of
all the existing sites of the $L2{}_{1}$ phase. It is reported by
Picozzi that some types of disorder might lead to additional states
at the Fermi level, thus reducing the spin polarization {[}5{]}. In
addition to the atomic order, the crystallographic orientation of
the Heusler thin film is important and may break the half metallicity.
Therefore, an annealing process is required to initiate the crystallization
and to induce the atomic ordering. It is thus of great interest to
investigate the annealing temperature ($T_{a}$) effects on the structural
and magnetic properties of CFA thin films. The purpose of this paper
is to use ferromagnetic resonance in microstrip line (MS-FMR) under
an in-plane and a out-of-plane applied magnetic field, combined to
vibrating sample magnetometry (VSM), in view of investigating the
correlation between structural and magnetic properties of CFA thin
films grown on Si or on MgO substrates and annealed at different temperatures.
A special attention will be given to the effect of $T_{a}$ on the
FMR linewidth and damping constant.

\section{Sample and experimental set up }

CFA films were grown on MgO(001) and thermally oxidized Si(001)/SiO2
substrates using a magnetron sputtering system with a base pressure
lower than $3\times10^{-9}$ Torr. Prior to the deposition of the
CFA films, a 4 nm thick MgO buffer layer was grown at room temperature
(RT) by RF sputtering from a MgO polycrystalline target under an Argon
pressure of 15 mTorr. Next, 10 nm and 50 nm thick CFA films, were
deposited at room temperature by DC sputtering under an Argon pressure
of 1 mTorr, at a rate of 0.1 nm/s. Finally, the CFA films were capped
with a MgO(4 nm)/Ta(4 nm) bilayer. After the growth of the stack,
the structures were ex-situ annealed at different temperatures (Ta
= 315$\lyxmathsym{\textdegree}C$, 415$\lyxmathsym{\textdegree}C$,
515$\lyxmathsym{\textdegree}C$ and 615$\lyxmathsym{\textdegree}C$)
during 15 minutes in vacuum. The structural properties of the samples
have been characterized by X-ray diffraction (XRD) using a four-circle
diffractometer. Their magnetic static and dynamic properties have
been studied by vibrating sample magnetometer (VSM) and microstrip
ferromagnetic resonance (MS-FMR) {[}6{]}, respectively.

\subsection{Structural properties }

XRD is a standard method for the characterization of the crystal growth
properties of thin films. Figure 1 shows the XRD $\theta-2\theta$
patterns for 50 nm thick CFA films grown on MgO substrates, annealed
at different temperatures and on Si substrates (grazing incidence
(GI) configuration) annealed at 615$\lyxmathsym{\textdegree}C$, respectively.
The XRD patterns show that, in addition to the (002) peak of the MgO
substrate, the CFA films (Fig. 1a) exhibit only two peaks which are
attributed to the (002) and (004) diffraction lines of CFA. The (004)
peak is expected for the A2 type structure, while the existence of
an additional (002) peak indicates a B2 type structure. Since the
ratio $I_{002}/I_{004}$ of the integrated intensities of the (002)
and of the (004) peaks increases versus $T_{a}$ (Fig. 1c), the chemical
order varies from the $B2$ phase towards the $A2$ phase when decreasing
Ta {[}7{]}. However, the I002/I004 ratio is the significantly below
the theoretical expected one for a fully $B2$ ordered crystal {[}8{]}.
This seems to indicate that, even after annealing the films at 615\textdegree{}C,
regions with different type of chemical order still coexist. Moreover,
using the well known Scherrer equation we determined a mean crystallite
size of around 10 nm for MgO grown films irrespective of annealing
temperature. This indicates that with annealing temperature we do
not have a grain growth of $B2$ but a crystallization of regions
with structural disorder and an improvement of the chemical ordering
towards the B2 phase, since both the intensities of (002) and (004)
peaks increase with annealing . In contrast, the GIXRD pattern of
CFA films grown on Si (Fig. 1b) clearly shows peaks corresponding
to CFA (022), (004) and (224) reflections, due to the lack of epitaxial
growth in the case of a Si substrate, which gives rise to a polycrystalline
structure. The diffraction patterns exhibit peaks ($hkl$) corresponding
only to the $h+k+l=4n$ (where $n$ is an integer) type reflections
which seems to indicate that the films show $A2$ structure with a
disorder among Co, Fe, and Al sites. However, a precise evaluation
of the chemical order is difficult to perform, due to the relative
low diffracted signal. Furthermore, pole figures (not shown here)
revealed that CFA grown on Si does not display any in-plane preferential
growth direction. In the case of the CFA films grown on MgO, $\phi$-scan
measurements (see inset of Figure 1a) allow for asserting an epitaxial
growth of the CFA films grown on MgO, within the investigated temperature
range for Ta, according to the expected CFA(001){[}110{]}//MgO(001){[}100{]}
epitaxial relation. Figure 1c shows the variations of the lattice
constant for increasing annealing temperatures for different samples.
The dashed lines represent the L21 ordered CFA bulk value {[}9{]}.
For the films deposited on thermally oxidized Si(001)/SiO2 substrates
the lattice constant was obtained by applying the Bragg equation to
the corresponding (022) peak. The lattice parameter is larger than
the reference value in the as prepared state and decreases with increasing
the annealing temperature for CFA for both substrates. The out-of-plane
and the in-plane lattice parameters of the CFA films grown on MgO
were evaluated using symmetric and asymmetric XRD scans of (004) and
(022) type reflections, respectively. From Fig. 1d one can observe
that the as-deposited film experiences a relatively strong tetragonal
distortion. As the annealing temperature increases the distortion
relaxes. In the case of the films annealed at temperatures higher
than 515$\lyxmathsym{\textdegree}C$ the value of out-of-plane is
essentially equal with the value of the in-plane lattice parameters.
This type of lattice parameter evolution was previously noticed in
CFA/MgO(001) samples {[}7{]} and was attributed the residual strain
associated with the growth method. Moreover, the lattice parameters
of CFA films of identical thicknesses (50 nm), grown on Si or on MgO
substrates are very close from each other in the investigated Ta range,
suggesting that the decrease of the lattice parameter is not due to
the chemical disorder, but most likely to relaxation of residual strain.

\begin{figure}
\includegraphics[bb=20bp 0bp 300bp 580bp,clip,width=8.5cm]{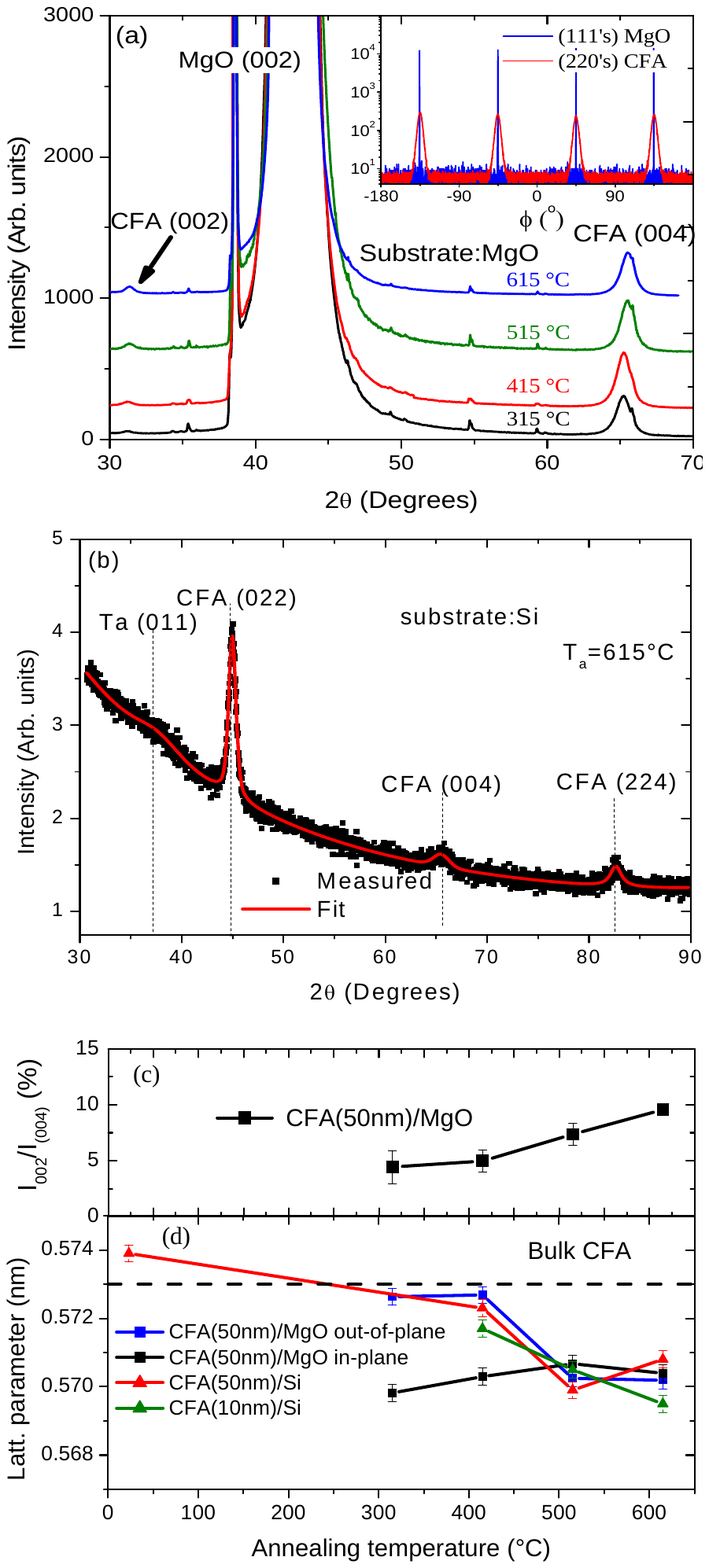}

\caption{: (Colour online) (a) X-ray $\theta-2\theta$ diffraction pattern
for the 50 nm CFA thick films grown on a MgO substrate annealed at
different temperature. The inset represents the $\phi$-scans of the
substrate and the CFA films annealed at 615$\lyxmathsym{\textdegree}C$.
(b) X-ray diffraction pattern measured in grazing incidence geometry
for the 50 nm CFA thick film grown on a Si substrate annealed at 615$\lyxmathsym{\textdegree}C$.
The symbols represent experimental data while the lines are the result
of the theoretical fit. The vertical dashed lines mark the positions
of the Ta(011) and CFA((022), (004), (224)) reflections. (c) Evolution
of the ratio of the integral intensities of the (002) and (004) Co2FeAl
peaks $I_{(002)/}I_{(004)}$ with respect to the annealing temperatures
of 50 nm thick Co$_{2}$FeAl films grown on MgO Substrates. (d) Evolution
of the lattice parameter, deduced from the position of the (022) and
of the (002) peaks, as function of the annealing temperature in 10
nm and 50 nm thick Co$_{2}$FeAl films grown on Si and CFA substrates. }

\end{figure}

\subsection{Magnetic properties }

All the measurements presented here have been made at room temperature
and analyzed using a model based on the following density of energy,
which was previously found to be appropriate to describe the properties
of Heusler films {[}10{]}:

\begin{multline}
E=-M_{S}H\left(\cos\left(\varphi_{M}-\varphi_{H}\right)\sin\theta_{M}\sin\theta_{H}+\cos\theta_{M}\cos\theta_{H}\right)\\
-\frac{1}{2}(1+\cos(2(\varphi_{M}-\varphi_{u}))K_{u}\sin^{2}\theta_{M}-\left(2\pi M_{S}^{2}-K_{\perp}\right)\sin^{2}\theta_{M}\\
-\frac{1}{8}(3+\cos4(\varphi_{M}-\varphi_{4}))K_{4}\sin^{4}\theta_{M}
\end{multline}

In the above expression, $\theta_{M}$ and $\varphi_{M}$ respectively
represent the out-of-plane and the in-plane (referring to the substrate
edges) angles defining the direction of the magnetization $M_{S}$.
$\varphi_{u}$ and $\varphi_{4}$ define define the angles between
an easy uniaxial planar axis or an easy planar fourfold axis, respectively,
with respect to this substrate edge. $K_{\mathit{u}}$, $K_{\mathit{4}}$
and $K_{\perp}$ are in-plane uniaxial, fourfold and out-of-plane
uniaxial anisotropy constants, respectively. We define $H{}_{\mathit{u}}=2K_{u}/M_{S}$
and $H_{4}=4K{}_{\mathit{4}}/M_{s}$ as the in-plane uniaxial and
the fourfold anisotropy fields respectively and we introduce the effective
magnetization $M_{eff}=H_{eff}/4\pi$ obtained by:
\[
4\pi M_{eff}=H_{eff}=4\pi M_{S}-\frac{2K_{\perp}}{M_{S}}=4\pi M_{S}-H_{\perp}
\]

The resonance expressions for the frequency of the uniform precession
mode and for the perpendicular standing spin waves (PSSW) modes assuming
in-plane or perpendicular applied magnetic fields are given in {[}6{]}.
The experimental results concerning the measured peak-to-peak FMR
linewidths $\Delta H^{PP}$ are analyzed in this work taking account
of both intrinsic and extrinsic mechanisms. As discussed in {[}6{]},
the observed magnetic field linewidth ($\Delta H^{PP}$) is analyzed
by considering Gilbert ($\Delta H^{Gi}$) {[}11{]}, inhomogeneities
($\Delta H^{inh}$) and two magnon scattering ($\Delta H^{2mag}$)
{[}12{]} contributions. This latter is given by {[}6{]}:
\begin{multline}
\Delta H^{2mag}=\Gamma_{0}+\Gamma_{2}\cos2(\varphi_{H}-\varphi_{2})+\\
\Gamma_{4}\cos4(\varphi_{H}-\varphi_{4})\arcsin\left(\frac{f}{\sqrt{f^{2}+f_{0}^{2}}+f_{0}}\right)
\end{multline}

with: $f_{0}=\gamma M_{eff}$. The expected fourfold symmetry induces
the $\Gamma_{0}$ and $\Gamma_{4}$ coefficients; the coefficient
$\Gamma_{2}$ is phenomenogically introduced. The total FMR linewidth
in our samples can be written as:
\begin{equation}
\Delta H^{PP}=\Delta H^{Gi}+(\Delta H^{mos}+\Delta H^{inh}+\Delta H^{2mag})
\end{equation}
The analysis of the variation of the resonance linewidth $\Delta H^{PP}$
versus the frequency and the in-plane field orientation allows for
evaluating $\alpha$, $\Delta\varphi_{H}$, $\Delta H^{inh}$, $\Gamma_{0}$,
$\Gamma_{2}$ (and $\varphi_{2}$) and $\Gamma_{4}$ (and $\varphi_{4}$
which, from symmetry considerations, is expected to have a $0\lyxmathsym{\textdegree}$
or $45\lyxmathsym{\textdegree}$ value, depending upon the chosen
sign of $\Gamma_{4}$).

\subsubsection{Static properties}

The easy axis VSM hysteresis loops were measured at different field
orientation and the magnetization at saturation has been extracted.
The magnetization at saturation of the three samples (not shown here)
increases slightly (10$\%$ of change) with increasing Ta, which indicates
an improved atomic ordering. A maximum value of 1029 emu/cm3 can be
achieved for a 50 nm CFA thick film grown on a MgO substrate and annealed
at 615$\lyxmathsym{\textdegree}C$. This value is higher, when compared
to a similar CFA film grown on a Si substrate, most probably due to
the higher crystalline quality and higher order degree. Figure 2 shows
typical hysteresis loops, along the easy axis, as function of the
annealing temperature for the 50 nm thick CFA films grown on Si and
on MgO substrates. For both substrates a clearly different behaviour
is observed between the samples annealed at low and at higher temperature.
The increase of the coercive field ($H_{C}$) with decreasing $T_{a}$
is an indication of the improving of the crystalline structure and
of the chemical order in the annealed samples. One should mention
that the angular dependence of $M_{r}/M_{S}$ (not shown here) of
annealed CFA films grown on Si showed a uniaxial anisotropy behaviour.
As the annealing temperature increases the sample quality and the
chemical order are enhanced and a unique uniaxial anisotropy is observed.
This is confirmed by the variation of the uniaxial anisotropy easy
axis direction with $T_{a}$ as shown below. For the 50 nm thick films
grown on a MgO substrate, the measured hysteresis loops revealed a
fourfold in-plane magnetic anisotropy with (110) easy and (100) hard
axis directions to which a weak uniaxial magnetic anisotropy is superimposed.
The intermixing of various chemical ordered phases or even possible
amorphous regions of CFA grown on MgO substrate, as revealed by XRD,
generates regions within the film with different magnetocrystalline
anisotropy. Therefore the measured hysteresis loops results in poor
squarness which gets improved as the annealing temperature increases
due to the enhancement of the disorder. Full square easy axis hysteresis
loop has been obtained for CFA thin films annealed at 750$\lyxmathsym{\textdegree}C$
(not shown here).

\begin{figure}
\includegraphics[bb=20bp 250bp 300bp 595bp,clip,width=8.5cm]{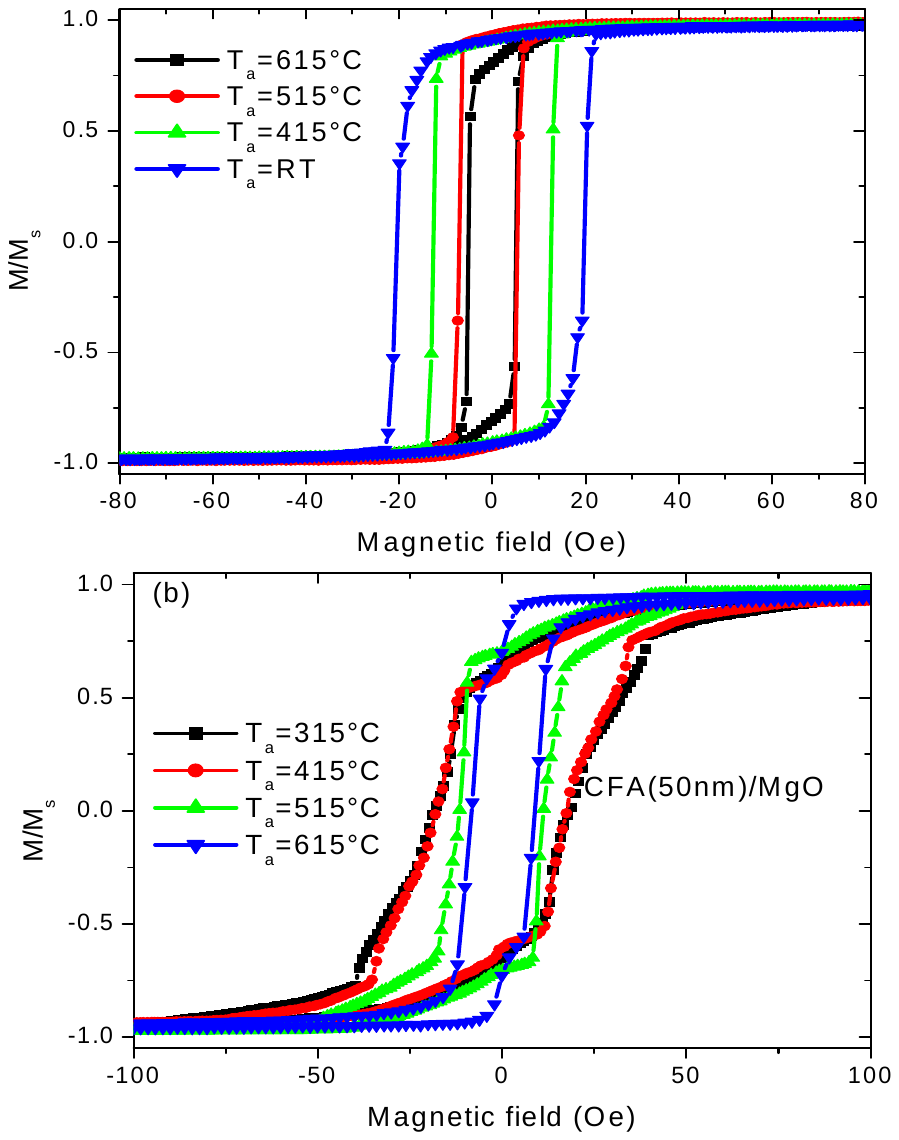}

\caption{(Colour on line) Hysteresis loops, along the easy axis, of the 50
nm thick CFA films annealed at different temperatures and grown on
(a) a Si substrate (b) a MgO Substrate. }
\end{figure}

\subsubsection{Dynamic properties}

\begin{figure}
\includegraphics[bb=20bp 420bp 280bp 595bp,clip,width=8.5cm]{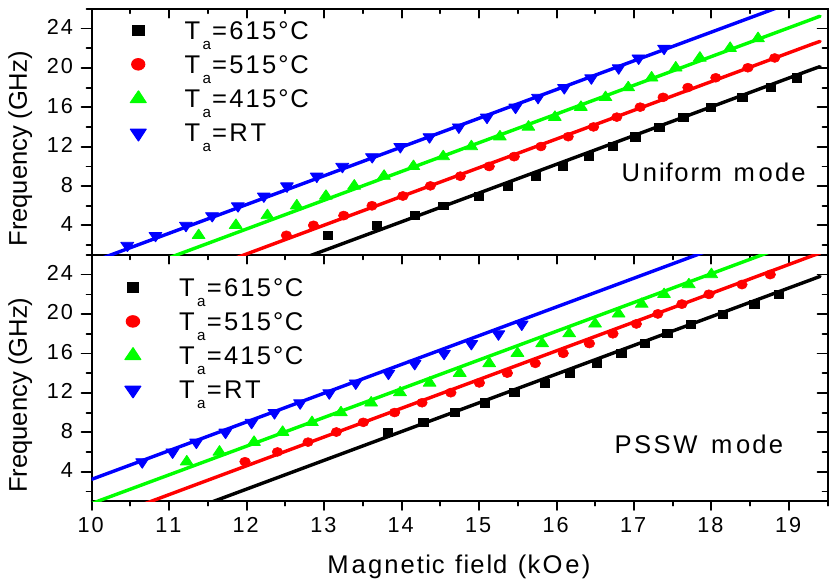}

\caption{(Colour on line) Variation of the frequencies of the uniform and of
the PSSW modes for 50 nm thick CFA films annealed at different temperatures.
Solid lines indicate the fit using the model described in the text. }
\end{figure}

\begin{figure}
\includegraphics[bb=20bp 235bp 300bp 595bp,clip,width=8.5cm]{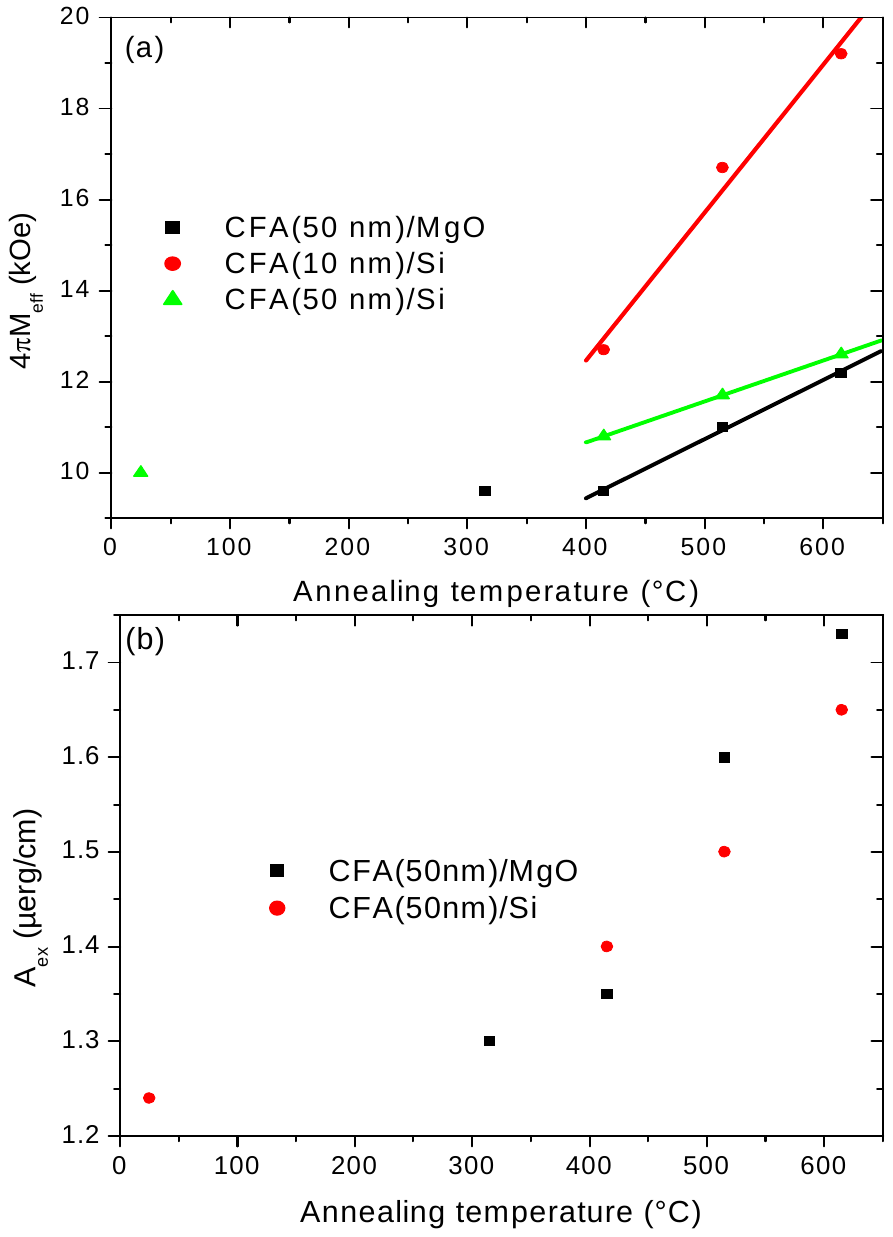}

\caption{(Colour on line) Variations of (a) the effective magnetization ($4\pi M_{eff}$)
and of the (b) exchange stiffness constant ($A_{ex}$), as function
of the annealing temperature, of 10 nm and 50 nm thick CFA films grown
on Si and on MgO substrates. The solid lines refer to the linear fit
in the range 400-650$\lyxmathsym{\textdegree}C$.}
\end{figure}

The uniform precession and the first perpendicular standing spin wave
(PSSW) modes have been observed in perpendicular and in-plane applied
field configurations for the 50 nm thick films while for the 10 nm
thick film, no PSSW mode is detected due to their high frequency over-passing
the available bandwidth (0-24 GHz). Typical perpendicular field dependences
of the resonance frequencies of the uniform and of the PSSW modes
are shown on figure 3 for the CFA film grown on Si. By fitting the
data in figure 3 to the model presented above, the gyromagnetic factor
($\gamma$), the exchange stiffness constant ($A_{ex}$) and the effective
magnetization ($4\pi M_{eff}$) are extracted. The fitted $\gamma/2\pi=29.2$
GHz/T is independent on $T_{a}$ while $A_{ex}$ (Fig. 4b) decreases
versus $T_{a}$, suggesting an enhancement of the chemical order when
the annealing temperature increases, presumably due to the enhancement
of the chemical order. A similar behaviour of the exchange stiffness
of Co$_{2}$FeAl$_{0.5}$Si$_{0.5}$ {[}13{]} with Ta has been reported
by Trudel et \textit{al}. The smaller $A_{ex}$ values of CFA films
grown on Si compared to those grown on MgO is another indication of
the best crystalline structure and chemical order of these latter.
\begin{figure}
\includegraphics[bb=20bp 235bp 300bp 595bp,clip,width=8.5cm]{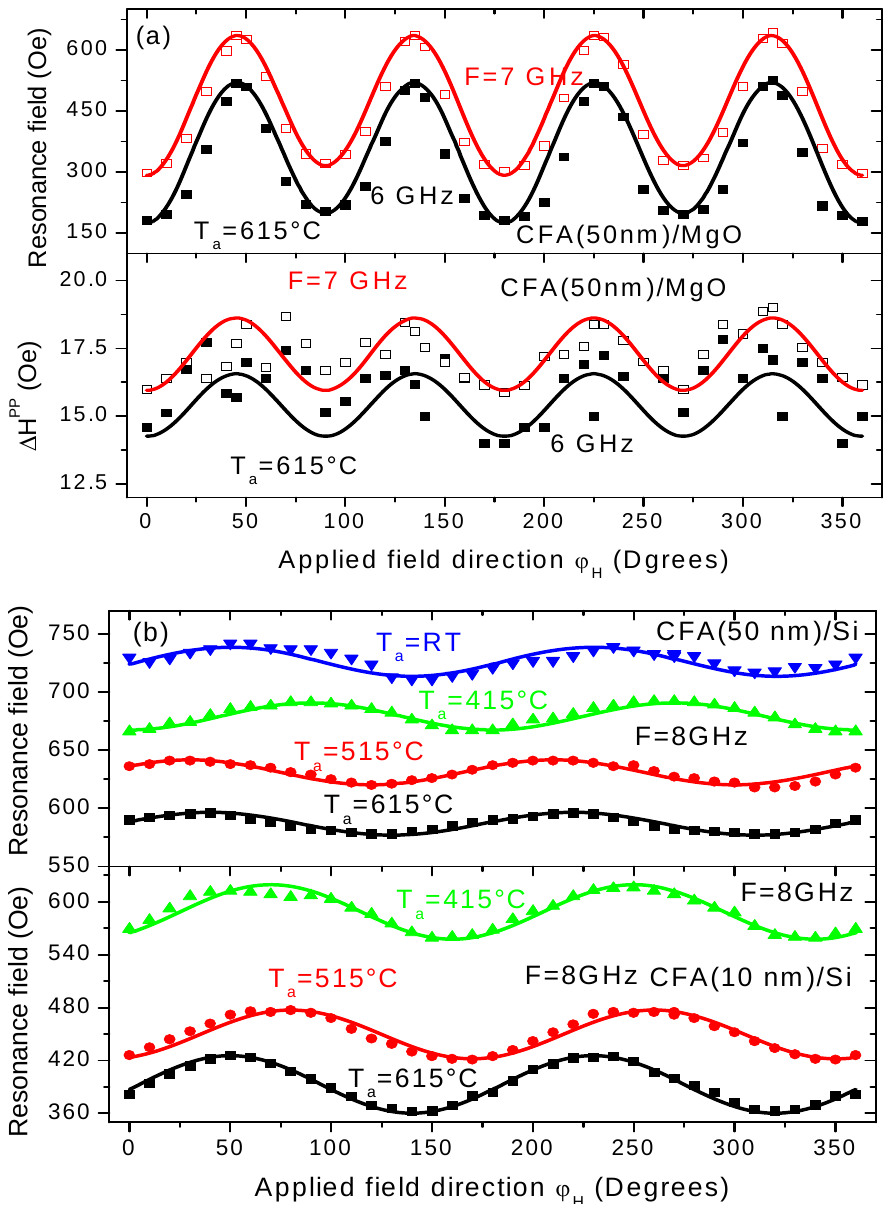}

\caption{(Colour online) Angular dependence of the resonance field and of the
peak to peak field FMR linewidth ($\Delta H^{PP}$) of 10 nm and 50
nm thick CFA thin films grown on (a) MgO and on (b) Si substrates.
The solid lines refer to the fit using the model described in the
text.}
\end{figure}

\begin{figure}
\includegraphics[bb=20bp 395bp 300bp 595bp,clip,width=8.5cm]{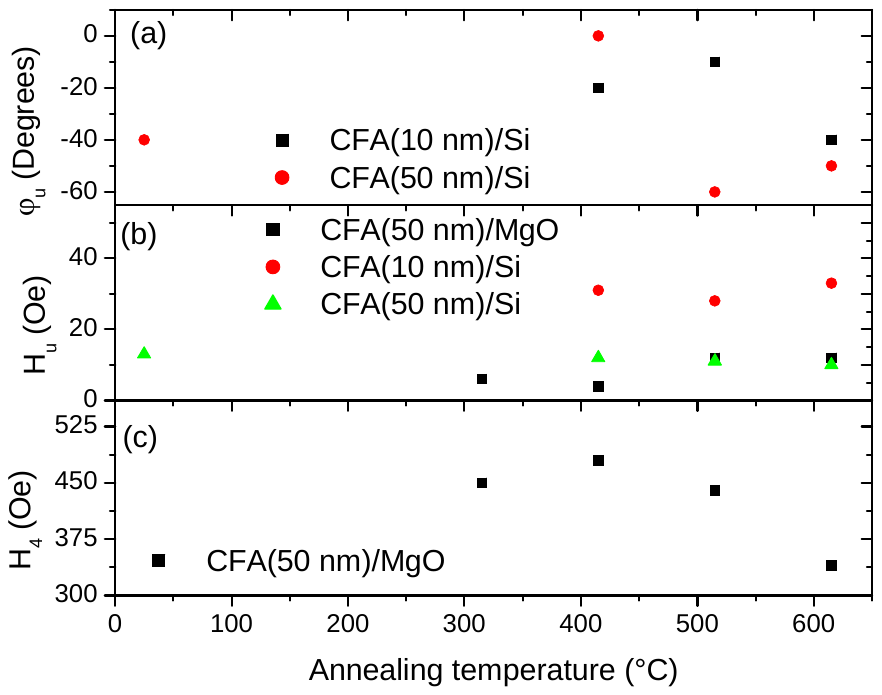}

\caption{(Colour online) Annealing temperature dependence of the (a) uniaxial
anisotropy easy axis direction (b) uniaxial ($H_{u}$) and of the
(c) fourfold anisotropy fields ($H_{4}$) of the 10 nm and 50 nm CFA
thick films grown on Si and MgO substrates.}
\end{figure}

Interestingly, the extracted effective magnetization from the MS-FMR
measurements, shown on figure 4a increases linearly with the annealing
temperature leading to negative perpendicular anisotropy, which tends
to favour the in-plane orientation. This effect is more pronounced
for the 10 nm thick film: the slope of this linear dependence, in
the range 415$\lyxmathsym{\textdegree}C$-615$\lyxmathsym{\textdegree}C$,
decreases from 32 Oe/$\lyxmathsym{\textdegree}C$ to 9 Oe/$\lyxmathsym{\textdegree}C$
respectively for the 10 nm and for the 50 nm thick CFA films grown
on Si. The 50 nm CFA thick film shows a larger slope (13 Oe/$\lyxmathsym{\textdegree}C$)
compared to the similar CFA film grown on Si. This effect originates
from the interface CFA/MgO, which is improved through the increasing
of the annealing temperature. Moreover, a linear variation of $4\pi M_{eff}$
as function of the thickness inverse of CFA films, annealed at $T_{a}=600\lyxmathsym{\textdegree}C$
and deposited on MgO {[}6{]} and on Si substrates, has been observed.
This thickness dependent anisotropy was also observed in structures
based on CoFeB/MgO {[}14{]}, but unlike CFA/MgO, the CoFeB/MgO interface
favours a perpendicular magnetization. Moreover, it is worth mentioning
in the case of Ta/CFA/MgO multilayers the surface anisotropy favours
also perpendicular magnetization {[}15{]}. Figures 5 shows the typical
MS-FMR angular dependence of the resonance field at 7 GHz and 8 GHz
driven frequency for three investigated CFA thick films annealed at
various Ta. For the 50 nm CFA thick film grown on MgO, the angular
dependence is only presented at $T_{a}=615\lyxmathsym{\textdegree}C$
for clarity. The MS-FMR measurements show that the angular behavior
of the resonance field is governed by a uniaxial anisotropy or a superposition
of uniaxial and fourfold anisotropies, respectively for films grown
on Si and on MgO substrates. The disappearance of the fourfold anisotropy
for CFA grown on Si is directly correlated to their in-plane polycrystalline
structure due to the amorphous SiO$_{2}$ layer. The uniaxial anisotropy
field ($H_{u}$), presented on Figure 6b, is unaffected by $T_{a}$
for all the samples. However, in contrast the CFA films grown on MgO,
the direction of the uniaxial anisotropy easy axis ($\varphi_{u}$),
determined with a 90$\lyxmathsym{\textdegree}$ precision since it
is referenced with respect to substrate edges, varies with Ta in the
case of CFA thin films grown on Si substrates as shown on Fig. 6a
, which complicates the identification of its origin. Therefore, a
completely satisfactory interpretation of the presence of Hu and of
its $T_{a}$ dependency is still missing. For the 50 nm CFA thick
film grown on MgO, the uniaxial and the fourfold anisotropies show
parallel easy axes which remain independent of $T_{a}$: this common
axis coincides with one of the substrate edges($\varphi_{u}=$$\varphi_{4}=0$)
and, consequently, with one of the $<110>$ crystallographic directions
of the CFA phase. The decrease of the fourfold anisotropy field ($H_{4}$)
as Ta increases (Fig. 6c) is an effect of the improving crystalline
structure and of the enhancement of the chemical order.

The FMR linewidth is a measure of the relaxation rate of the magnetization
and is related to the magnetic damping. This linewidth is caused by
two mechanisms: the intrinsic damping of the magnetization and extrinsic
contributions {[}16{]} (such as two magnons scattering, mosaicity,\ldots{}).
The angular and frequency dependences of the FMR linewidth provide
information about these magnetic damping mechanisms. Therefore, the
field peak to peak FMR linewidth, defined as the field difference
between the extrema of the sweep-field measured FMR spectra, has been
investigated as function of the annealing temperature.

In figure 5a, the FMR peak to peak linewidth ($\Delta H^{PP}$) is
plotted as a function of the field angle $\varphi_{H}$, using 6 GHz
and 7 GHz driving frequencies, for the 50 nm CFA films grown on a
MgO substrate and annealed at 615\textdegree{}C. In the CFA samples
grown on MgO, the $\Delta H^{PP}$ angular variation shows a perfect
fourfold symmetry (in agreement with the variation of the resonance
position) while it shows a uniaxial behaviour in the case of CFA grown
on Si. Such behavior is characteristic of a two magnons scattering
contribution. This effect is correlated to the presence of defects
preferentially oriented
\begin{figure}[H]
\includegraphics[bb=20bp 20bp 300bp 595bp,clip,width=8.5cm]{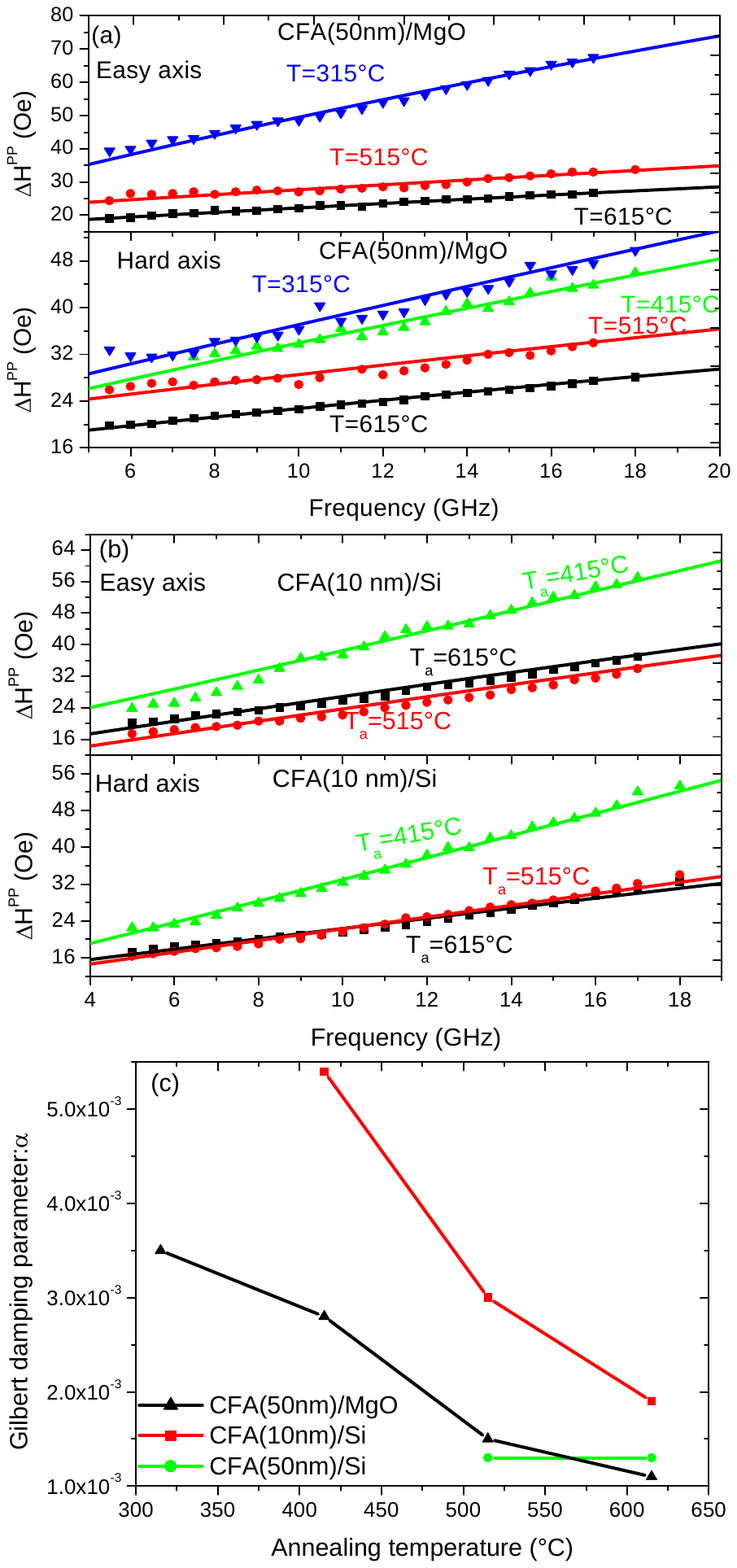}

\caption{(Colour online) Frequency dependence of the easy and of the hard axes
peak to peak field FMR linewidth ($\Delta H^{PP}$) for (a) 50 nm
CFA thick films grown on MgO and for (b) 10 nm CFA thick films grown
on Si. The solid lines refer to the fit using the model described
in the text. (c) Annealing temperature dependence of the Gilbert damping
parameter of 10 nm and 50 nm CFA thin films grown on Si and MgO substrates. }
\end{figure}

\begin{table*}[!t]
\begin{singlespace}
\noindent %
\begin{tabular}{>{\centering}p{1.2cm}>{\centering}p{1.5cm}>{\centering}p{1.2cm}>{\centering}m{1.24cm}>{\centering}p{2cm}>{\centering}p{2cm}>{\centering}p{2cm}>{\centering}p{2cm}>{\centering}p{1.2cm}}
\hline
$d$ (nm)  & $\Gamma_{0}$ & $\Gamma_{1}$  & $\Gamma_{2}$ & $\varphi_{2}$ (deg.)  & $\varphi_{4}$ (deg.) & $\Delta H^{inh}$ (Oe) & $\alpha$ & $T_{a}$ ($\lyxmathsym{\textdegree}C$)\tabularnewline
\hline
\hline
\multicolumn{9}{c}{MgO substrate}\tabularnewline
\hline
\hline
50 & 11 & np & -1 & np & 0 & 15.2 & $1.1\times10^{-3}$ & 615\tabularnewline
50 & 8 & np & -1 & np & 0 & 20 & $1.5\times10^{-3}$ & 515\tabularnewline
50 & 29 & np & 17 & np & 0 & 20 & $3.5\times10^{-3}$ & 315\tabularnewline
\hline
\hline
\multicolumn{9}{c}{Si substrate}\tabularnewline
\hline
\hline
10 & 35 & 13 & np & -40 & np & 11 & $1.9\times10^{-3}$ & 615\tabularnewline
10 & 12 & -8 & np & -10 & np & 8.5 & $3\times10^{-3}$ & 515\tabularnewline
10 & 31.5 & -7.5 & np & -20 & np & 1 & $5.4\times10^{-3}$ & 315\tabularnewline
\hline
\end{tabular}
\end{singlespace}

\caption{{\small Magnetic parameters obtained from the best fits to our experimental
FMR results with the above-mentioned model. np: not pertinent.}}
\end{table*}
 along specific crystallographic directions, thus leading to an anisotropic
damping. All the other samples show a qualitatively similar behaviour
to one of the samples presented here. The positions of the extrema
depend on the sample. The observed pronounced anisotropy of the linewidth
cannot be due to the Gilbert damping contribution, which is expected
to be isotropic, and must be due to additional extrinsic damping mechanisms.

Figure 7 presents the frequency dependence of FMR linewidth, for an
applied field parallel to the easy and the hard axes, of 50 nm and
10 nm thick CFA films annealed at various temperatures and grown respectively
on MgO and Si substrates. It shows that the linewidth decreases with
increasing $T_{a}$ due to the enhancement of the chemical order.
The observed angular and frequency dependences of the field linewidth
have been analyzed conjointly by considering intrinsic (Gilbert),
two magnons scattering and inhomogeneities contributions, using the
similar method described in {[}6{]} and expression (3). Consequently,
$\Gamma_{0}$, $\Gamma_{2}$,$\Gamma_{4}$, $\varphi_{2}$, $\varphi_{4}$
modelling the two magnon contribution, obtained from the best fit
of the FMR data are listed in Table I, which also contains the parameters
describing the damping effects and of the inhomogeneity. Figure 7c
shows the Ta dependence of the Gilbert damping constant ($\varphi_{4}$).
The CFA films grown on MgO present a smaller damping parameter compared
to those grown on Si. The 50 nm thick CFA film, annealed at 615$\lyxmathsym{\textdegree}C$,
shows a very low $\alpha$, equal to 0.0011, which is comparable to
that of the epitaxial Fe0.73V0.27, which is considered as the ferromagnetic
metal having the lowest $\alpha$ {[}17{]}. However, the 10 nm thick
sample shows relatively large values of $\alpha$ which decrease with
increasing $T_{a}$. The reason for both these larger values for thinner
films and for the decrease of $\alpha$ with increasing $T_{a}$,
is most probably due to increase of the chemical order degree {[}18{]}.

\section{Conclusion}

Co$_{2}$FeAl films with thicknesses of 10 nm and of 50 nm were prepared
by sputtering on Si(001)/SiO2 and MgO(001) substrates and annealed
at various temperatures. They show a polycrystalline structure and
an epitaxial growth when Si and MgO substrates are used, respectively.
The chemical order changes from $A2$ structure to $B2$ when increasing
the annealing temperature ($T_{a}$). The microstrip ferromagnetic
resonance (MS-FMR) has been used to study the dynamic properties.
The MS-FMR has been adjusted to a model allowing for the determination
of the most relevant parameters. The in-plane uniaxial anisotropy
field, present in all the samples, is unaffected by $T_{a}$ and the
fourfold anisotropy field, observed in the sample grown on MgO, decreases
when increasing the annealing temperature. The presence of this fourfold
anisotropy is directly correlated to the crystalline structure of
CFA grown on MgO. The effective magnetization increases drastically
with $T_{a}$, due to the enhancement of the CFA/MgO interface quality.
The angular and frequency dependences of the FMR linewidth, which
decreases with increasing annealing temperature, are governed by two
magnons scattering and by a Gilbert damping which decreases with the
increasing annealing temperature due to the chemical disorder.
\begin{acknowledgments}
This work was partially supported by POS CCE Project ID.574, code
SMIS-CSNR 12467 and \textquotedbl{}SPINTAIL\textquotedbl{} PN-II-ID-PCE-2012-4-0315. \end{acknowledgments}


\begin{thebibliography}{10}
\bibitem{key-1}{[}1{]} H. C. Kandpal, G. H. Fecher, and C. Felser,
J. Phys. D 40, 1507 (2007)

\bibitem{key-2}{[}2{]} R. A. de Groot, F. M. Mueller, P. G. van Engen
and K. H. J. Buschow, Phys. Rev. Lett. 50, 2024 (1983)

\bibitem{key-5}{[}3{]} W. H. Wang, H. Sukegawa, and K. Inomata, Phys.
Rev. B 82, 092402 (2010)

\bibitem{key-3}{[}4{]} W. H. Wang, H. Sukegawa, R. Shan, S. Mitani,
and K. Inomata, Appl. Phys. Lett. 95, 182502 (2009).

\bibitem{key-4}{[}5{]} S. Picozzi, A. Continenza and A. J. Freeman,
Phys. Rev. B 69, 094423 (2004)

\bibitem{key-6}{[}6{]} M. Belmeguenai, H. Tuzcuoglu, M. S. Gabor,
T. Petrisor Jr. C. Tiusan, D. Berling, F. Zighem, T. Chauveau, S.
M. Chérif and P. Moch, Phys. Rev. B, 87, 184431 (2013)

\bibitem{key-7}{[}7{]} M. S. Gabor, T. Petrisor Jr., and C. Tiusan,
M. Hehn and T. Petrisor, Phys. Rev. B 84, 134413 (2011)

\bibitem{key-8}{[}8{]} K. Inomata, N. Ikeda, N.Tezuka, R. Goto, S.
Sugimoto, M. Wojcik and E. Jedryka, Sci. Technol. Adv. Mater. 9, 014101,
(2008)

\bibitem{key-12}{[}9{]} H. J. Elmers, S. Wurmehl, G. H. Fecher, G.
Jakob, C. Felser and G. Schönhense, Appl. Phys. A 79, 557 (2004)

\bibitem{key-11}{[}10{]} M. Belmeguenai, F. Zighem, Y. Roussigné,
S. M. Chérif, P. Moch, K. Westerholt, G. Woltersdorf and G. Bayreuther
Phys. Rev. B79, 024419 (2009)

\bibitem{key-10}{[}11{]} B. Heinrich and J. F. Cochran, J. Appl.
Phys. 57, 3690 (1985).

\bibitem{key-9}{[}12{]} A. K. Srivastava, M. J. Hurben, M. A. Wittenauer,
P. Kabos, C. E. Patton, R. Ramesh, P. C. Dorsey and D. B. Chrisey,
J. Appl. Phys.85, 7838 (2009)

\bibitem{key-13}{[}13{]} S. Trudel, G. Wolf, J. Hamrle, B. Hillebrands,
P. Klaer, M. Kallmayer,H. J. Elmers, H. Sukegawa, W. Wang, and K.
Inomata, Pysc. Rev. B 83, 104412 (2011)

\bibitem{key-14}{[}14{]} S. Ikeda, K. Miura, H. Yamamoto, K. Mizunuma,
H. D. Gan. M. Endo, S. Kanai, J. Hayakawa, F. Matsukura and H. Ohno,
Nature Mater. 9, 721 (2010)

\bibitem{key-15}{[}15{]} M. S. Gabor, T. Petrisor Jr., C. Tiusan
and T. Petrisor, J. Appl. Phys. 114, 063905 (2013)

\bibitem{key-16}{[}16{]} Kh. Zakeri, J. Lindner, I. Barsukov, R.
Meckenstock, M. Farle, U. von Hörsten, H. Wende, W. Keune, J. Rocker,
S. S. Kalarickal, K. Lenz, W. Kuch and K. Baberschke, Phys. Rev. B
76, 104416 (2007)

\bibitem{key-17}{[}17{]} C. Scheck, L. Cheng, I. Barsukov, Z. Frait
and W. E. Bailey, Phys. Rev. Lett. 98, 117601 (2007)

\bibitem{key-18}{[}18{]} S. Mizukami, D. Watanabe, M. Oogane, Y.
Ando, Y. Miura, M. Shirai and T. Miyazaki, J. Appl. Phys.105, 07D306
(2009) \end{thebibliography}
\end{document}